\documentclass{article}

\newcommand{\ba}{\begin{eqnarray}}
\newcommand{\ea}{\end{eqnarray}}
\newcommand{\be}{\begin{equation}}
\newcommand{\ee}{\end{equation}}
\newcommand{\bib}{\bibitem}

\begin{document}

\thispagestyle{empty}
$\phantom{x}$\vskip 0.618cm



\begin{center}
{\huge  Dual Projection and Selfduality in Three Dimensions.}\\
\vspace{1.5cm}
{\Large Rabin Banerjee${}^{b}$ and Clovis Wotzasek${}^{a,b}$}\\
\vspace{1.5cm}
{\em ${}^{a}$Instituto de F\'\i sica\\Universidade
Federal do Rio de Janeiro\\21945, Rio de Janeiro, Brazil\\}

{\em ${}^{b}$S. N. Bose National Centre for Basic Sciences,\\
Block JD, Sector III, Salt Lake, Calcutta 700091, India.\\}
\end{center}

\begin{abstract}
\noindent We discuss the notion of duality and selfduality in the
context of the dual projection
operation that creates an internal space of potentials.
Contrary to the prevailing algebraic or group theoretical methods, 
this technique is applicable to both even and odd dimensions. 
The role of parity in the kernel of the Gauss law to determine the
dimensional dependence
is clarified.
We derive the appropriate invariant actions, discuss the symmetry
groups and their proper generators.
In particular, the novel concept of duality symmetry and 
selfduality in Maxwell theory in (2+1) dimensions is analysed
in details.
The corresponding action is a 3D version of the familiar duality
symmetric electromagnetic
theory in 4D. Finally, the duality symmetric actions in the different dimensions
constructed here manifest both
the $SO(2)$ and $Z_2$ symmetries, contrary to conventional results.

\end{abstract}
\newpage

\section{Introduction}

It is often claimed that the duality operation is only defined in even dimensional
spacetimes and that selfduality is further restricted to twice-odd spacetime
dimensional theories\cite{PST1}.
The purpose of this paper is to extend the notion of both duality symmetry
as well as selfduality to the odd (2+1)
dimensional spacetime in the electromagnetic context. Naturally, the conventional
duality symmetry in even dimensions is also contained in our approach.
To achieve this we introduce an
alternative definition for the duality operation which is valid
in any dimensions.
Specifically we analyse and explore the impact of the Gauss law kernel's parity
and the dual projection procedure over the duality operation of the Maxwell
theory in different spacetime dimensions.  The crucial issue
concerns the spacetime dependence of the parity property of a
generalized curl involved in the resolution of the Gauss law.
We show that this property is decisive in
determining the proper actions and the corresponding group of symmetry for
each specific dimension.

 The role of the duality operation in the investigation of concrete
physical systems in different areas is by now well recognized\cite{AG}.  This is a symmetry
transformation that is fundamental for investigations in arenas as distinct as
quantum field theory, statistical mechanics and string theory. Duality is a general
concept relating physical quantities in different regions of the parameter space. 
It relates a model in a strong coupling regime to its dual version working in a
weak coupling regime, providing valuable information in the study of strongly
interacting models.  The selfduality present in D=4k-2 dimensions has attracted
much attention because it seems to play an important role in many theoretical
models\cite{EK}.

The electromagnetic duality transformation is defined by the Hodge-star operation
that involves multiplication by the appropriate $\epsilon$ symbol.
Consider a general (n-1)-form and its field strength

\be
\label{a05}
F_{k_1 \cdots k_n} \equiv \partial_{[k_n} A_{k_1 \cdots k_{n-1}]}  .
\ee
The dual field is then defined as,

\be
\label{a10}
\mbox{}^* F^{k_1\cdots k_n} = \frac 1{n!} \; \epsilon^{k_1\cdots k_{2n}} F_{k_{n+1}\cdots k_{2n}}
\ee
Note that only in $2n$ dimensions will the $n$-form field be of the same rank as its dual.
The action, field equation and Bianchi identity for a source free field are

\ba
\label{a15}
S&=& -c_n \int d^{2n} x F_{k_1\cdots k_n }F^{k_1\cdots k_n} \nonumber\\
0&=& \partial_{k_1}F^{k_1\cdots k_n} \nonumber\\
0&=& \partial_{k_1}\mbox{}^*F^{k_1\cdots k_n} 
\ea
where $c_1= 1/2$, $c_2=1/4$, etc.  The field equation and the Bianchi identity
are of the same form so that the duality transformation $F\leftrightarrow \mbox{}^* F$
is a symmetry at the level of these equations but is not present at the level of the action.
The dependence with dimensionality appears to be crucial.  The duality symmetry
is characterized by a one-parameter continous SO(2) group for D=4k ($k\in Z_+$)
while for the D=4k-2 case it is described by a discrete group with just two elements. 
Most of the analysis of this dimensional dependence take the algebraic point of view
where the distinction among different dimensions is manifest by the double
dualization operation following from the identities,

\be
\label{a20}
\mbox{}^{**}F = \cases{F,&if $D=4k-2$\cr
	-F,&if $D=4k$.\cr}
\ee
It was then shown that the duality groups $G$ preserving the form of the action
were subgroups of those preserving equations of motion and Bianchi identities,
obtained by taking the intersection of the former with the group O(2), the
symmetry group for the energy-momentum tensor\cite{DGHT}.

Most of the discussion about duality transformations as a symmetry for the
actions and the existence of self-duality are based on
these concepts.
We feel that this algebraic viewpoint is rather restrictive.  It is only
defined for even dimensional spacetimes leading to separate consequences
regarding the duality groups and actions.  The usual lore is that only 
the 4D Maxwell theory and its 4k extensions possess duality as a symmetry
but selfduality would only be definable in D=4k-2.  On the other hand for
the 2D scalar field theory and its 4k-2 extensions duality is not even a well
defined concept.
A solution for these problems came with the recognition of an internal
structure in the space of potentials\cite{Zwanziger, BW}.
The internal space effectively 
unifies the selfduality concept in the different $4k-2$
and $4k$ dimensions. The dual field is now defined to include an internal
index $(\alpha, \beta)$ and an extended dualization is defined as,
\begin{eqnarray}
\label{ssi2}
\tilde F^\alpha &=&\epsilon^{\alpha\beta}\mbox{}^{*}F^\beta 
\,\,\,;\,\,\,D=4k\nonumber\\
\tilde F^\alpha &=&\sigma_1^{\alpha\beta}\mbox{}^{*}F^\beta 
\,\,\,;\,\,\,D=4k-2
\end{eqnarray}
where $\sigma_k$ are the usual Pauli matrices and 
$\epsilon_{\alpha\beta}$ is the fully
antisymmetric $2\times 2$ matrix with $\epsilon_{12} =1$. Now, 
irrespective of the
dimensionality, the double dual operation yields,
\begin{equation}
\label{ssi3}
\tilde{\tilde F} = F
\end{equation}
which generalises (\ref{a20}) . 
Self and anti-self dual solutions are now well defined in
all even $D=2k$ dimensions. 

With the above background, it is useful to put our work in a proper perspective
by observing the following:

\begin{itemize}

\item The procedure of dual projection developed here 
is basically analogous to a canonical transformation, except that
the former is performed at the level of the actions while the latter, as is well 
known, is at the hamiltonian level. Since throughout the paper only first order 
actions will be considered the equivalence between the dual projection and the
canonical transformations becomes manifest\cite{BG}.

\item The Maxwell action in any even dimension is decomposed, by the
dual projection method, into two pieces, one of which carries the $SO(2)$
symmetry while the other has the $Z_2$ symmetry. By specialising to $4k$ or
$4k-2$ dimensions, we find that one of the terms in the action becomes a total
derivative which can be ignored. In this way
the conventional results
of duality symmetry characterising the $SO(2)$ group for $D=4k$ and the $Z_2$ group for
$D=4k-2$ dimensions are reproduced.

\item The dual projection in D=4k-2 dimensions leads to a diagonal form of
the actions with two pieces manifesting the opposite chiralities. It is then
possible to impose a chiral constraint to eliminate either of the pieces.
What remains is the action for a chiral boson. This generalises the usual
construction of chiral bosons\cite{FJ} to higher dimensions.

\item The case D=2 seems to be a special point.  Although it qualifies
as a member of the D=4k-2 group, it will be shown  that it 
allows for the realization of both D=4k-2 and D=4k constructions.

\item By passing to the momentum space, it is possible to derive results
analogous to the D=2 case for higher dimensions. In other words there is a
duality transformation
among the Fourier modes showing the complementary nature of the symmetries:
the $SO(2)$ symmetry for $D=4k-2$ dimensions and the $Z_2$ symmetry for the $D=4k$
dimensions. It might be mentioned that this nature of duality symmetry was earlier
shown by us\cite{CW} using different methods.

\item The algebraic analysis leaves out the possibility
of obtaining duality symmetric electromagnetic Maxwell theory in odd dimensions. For
the special case of three dimensions this will
be achieved here.

\end{itemize}

In the next section we shall show that the use of the two key
concepts; namely, the dimensional dependence of the
parity property of the generalized rotation operator involved in the resolution
of the Gauss law and the dual projection method,
reproduce the known results of the algebraic analysis,
clarifying their physical origins.
In the third section we discuss some special instances like the
two-dimensional case that possess both $SO(2)$ and $Z_2$ representations.
In the fourth section, which contains the centeral result of this paper,
attention is given to the D=(2+1) Maxwell theory that is studied in
full details. We disclose the presence of an internal space of potentials
where duality is realized as a $SO(2)$ rotation and also as a discrete 
$Z_2$ symmetry. The special equivalence
between Maxwell theory and the scalar field via Hodge-dualization is discussed
from our approach.
The
last section is reserved to a discussion of our conclusions and perpectives.

\section{Parity and Dual Projection in Even Dimensional Spacetimes}

The main argument of this report is the dimensional dependence of the parity
property of the generalized rotation operator and a canonical transformation that we call
dual projection.
A systematic derivation of selfdual actions for two and four dimensional cases
was proposed in \cite{BW} using the dual projection procedure. Here we generalise the
method to inclued all even dimensions. Subsequently these ideas will be exploited to
discuss the consequences in a three dimensional theory.
The generalized rotation operator, which is basically a functional curl, is defined as,

\be
\label{a30}
(\epsilon\partial) \equiv \epsilon_{k_1 k_2 \cdots k_{D-1}}\partial_{k_{D-1}}
\ee
Clearly the dimensional dependence of this operator's parity is given by,

\be
\label{a40}
{\cal P}(\epsilon\partial) = \cases{+ 1,&if $D=4k$\cr
	-1,&if $D=4k-2$,\cr}
\ee
where parity is defined as 

\be
\label{a50}
\int \Phi(\epsilon\partial\Psi) = {\cal P}(\epsilon\partial)\int \Psi(\epsilon\partial\Phi)
\ee
The consequence of this property is best appeciated after a dual projection
of the action. First, the
theory is reduced to its first-order form as,

\be
\label{a60}
S= \int d^D x \left[\pi \cdot \dot A - \frac 12 \pi\cdot\pi -
\frac 12 B\cdot B +A_0(\partial \cdot \pi)\right]
\ee
where we used the notation (anti symmetrisation is implied by the brackets)
\be
\label{a70}
\Phi\cdot\Psi \equiv \Phi_{[k_1 k_2 \cdots k_{D-1}]}\Psi_{[k_1 k_2 \cdots k_{D-1}]}
\ee
and defined the magnetic field as

\be
\label{a80}
B = (\epsilon\partial)\cdot A
\ee

In the four dimensional case, $B_k=\epsilon_{kmn}\partial_m A_k$ is a
three-vector while in three dimensions, the magnetic field is a scalar,
$B=\epsilon_{km}\partial_m A_k$.  Clearly the two-dimensional instance
represents a special situation due to the absence of a Gauss law and
will be treated separately in the next section. There does not seem
to exist any difficulty in dimensions $D > 4$.
Note that $A_0$ in (\ref{a60}) generically denotes the multiplier in any even dimension,
that enforces the Gauss constraint. For example, it is just $A_0$ in four
dimensions while it is $A_{0i}$ in six dimensions and so on.    

The important point to observe is that the Gauss constraint is
trivially solved, in any dimension, using the generalized curl (\ref{a30}),

\be
\label{a90}
\pi=(\epsilon\partial)\cdot\phi
\ee
where $\phi$ is a ($\frac D2 - 1$)-form potential. For instance,
in D=4 and D=6 which are generic for D=4k and D=4k-2, this solution reads

\ba
\label{a100}
\pi_k &=& \epsilon_{kmn}\partial_m \phi_n\nonumber\\
\pi_{km} &=& \epsilon_{kmnpq}\partial_n \phi_{pq}
\ea
The next step is to perform the canonical transformations,

\ba
\label{a110}
A &=& \Phi^{(+)} + \Phi^{(-)}\nonumber\\
\pi &=& \eta (\epsilon\partial)\cdot(\Phi^{(+)} - \Phi^{(-)})
\ea
with $\eta = \pm 1$ defining the signature of the operation.  
The effect of the dual projection procedure into the first-order Maxwell
action is the creation of an internal space of potentials in which the
duality symmetry is local and manifest.  In terms of the internal space
potentials $\Phi^{(+)}$ and $\Phi^{(-)}$ the action now reads,

\be
\label{a120}
S= \int d^D x\left\{ \eta\left[\dot\Phi^{(\alpha)}\sigma_3^{\alpha\beta}B^{(\beta)} +
\dot\Phi^{(\alpha)}\epsilon^{\alpha\beta}B^{(\beta)}\right]- B^{(\beta)} B^{(\beta)}\right\}
\ee
where $B^{(\beta)}=(\epsilon\partial\cdot\Phi^{(\beta)})$
and $\sigma_3^{(\alpha\beta)}$ and $\sigma_2^{(\alpha\beta)}=i\,
\epsilon^{(\alpha\beta)}$ are the $2\times 2$ Pauli matrices.
Notice that while the hamiltonian sector of the first-order
action is unique, the symplectic sector is composed by two
distinct parts with separate consequences.
We can now appreciate the impact of the dimensionality over
the symplectic structure of (\ref{a120}) and the role of the
parity in selecting the proper action and the corresponding
duality group.  Parity (or dimensionality) has no
influence over the hamiltonian since it only involves quadratic
forms.  For twice odd dimensions the second term of the symplectic
sector is a total derivative and may be discarded.  The remaining
piece diagonalizes the action providing a generalization of the
two-dimensional action describing chiral bosons\cite{FJ}.  
The $Z_2$ property is manifest by the interchange between the internal
space potentials $\Phi^{(\pm)}\rightleftharpoons\Phi^{(\mp)}$ mapping
one chirality into the other.

For twice even dimensions, on the other hand, it is the first term that
becomes a total derivative. The action does not diagonalize but
presents an explicit one-parameter continous SO(2) symmetry. 
In the D=4 case this action corresponds to duality symmetric
Maxwell theory quoted in the literature \cite{Zwanziger, BW}.
The important point to stress  is that the derivative operator
involved in the dual projection has been determined by the solution
of the Gauss constraint.  This automactically fixes the dependence
of parity with dimensionality and explains its effect over the
electromagnetic actions and duality groups.  Incidentally, observe
that due to its intrinsic diagonal form, the phase space of the
chiral boson solution (D=4k-2) may be reduced if we impose a chiral
constraint as,

\be
\label{a130}
\pi = \pm (\epsilon\partial)\cdot A
\ee
Each of these constraints eliminates one of the (internal) chiral potentials
thereby leading to an action for chiral bosons. These are the generalisations
of the usual actions for chiral bosons in two
dimensions\cite{FJ}. However,
the same situation cannot be reached in the
twice-even instance due to the special form of the sympletic sector
(the hamiltonian poses no obstruction to reduction in either case). 
In the reduced phase-space the remaining chiral boson carries a representation
for half the number of degrees of freedom of the original system.  On the other
hand, the duality symmetric system mantains the phase space structure intact. 
Therefore, this system should not be considered as the 4D analog of the 2D chiral boson.
Although, due to the possibility of two distinct signatures in the dual projection,
there exist either a self-dual or anti self-dual decomposition (but not simultaneously),
we believe that this situation should not be confused with the distinct chiralities that
appear (simultaneously) in the dual projection of D=4k-2 dimensional systems.

We have therefore reproduced completely the results known from the algebraic approach
plus presented a derivation of the appropriate actions displaying the internal potentials
for each case.  Also worth of mention is the fact that there are two and not one self-dual
action, labelled by the signature $\eta$ of the dual projection, describing opposite
aspect of the self-duality symmetry.  
The most useful and striking feature in this dual projection procedure is that it is
not based on evidently even dimensional concepts and may be extended to the odd
dimensional situation.

\section{Special Examples}

In this section we discuss some special dimensions and situations.  The 4D Maxwell
example deserves detailed analysis since it is the paradigm of the duality symmetry.
The other example is the (1+1) dimensional scalar theory. The absence of a
gauss constraint leads to a crucial change from the
Maxwell case.

\subsection{The Electromagnetic Duality}

Exploiting the ideas elaborated in the previous sections, it is straightforward
to implement the selfduality projection in the electromagnetic theory. Let us start with
the usual Maxwell action,
\be
S =-\frac{1}{4}\int d^4x\; F_{\mu\nu}F^{\mu\nu}
\label{ssm10}
\ee
which is expressed in terms of the electric and magnetic fields
as,
\be
S= \frac{1}{2}\int d^4x\;\Big( E_k^2- B_k^2\Big)
\label{ssm20}
\ee
where,
\ba
E_i&=&-F_{0i}=-\partial_0 A_i+\partial_i A_0\nonumber\\
B_i&=&\epsilon_{ijk}\partial_j A_k
\label{ssm30}
\ea
The following duality transformation,
\be
E_k\rightarrow \mp B_k\,\,\,\,;\,\,\,\, B_k\rightarrow \pm  E_k
\label{ssm40}
\ee
is known to preserve the invariance of the full set comprising
Maxwell's equations and the Bianchi identities although the action
changes its signature.  The Maxwell Lagrangean
is next recast in a symmetrised first order form that displays an Sp(2,R)
symmetry when we treat ($P_k\, ,\, A_k$) as a doublet,
\be
{\cal L}=\frac{1}{2}\Big( P_k\dot{ A_k}-\dot{P}_k A_k\Big)
-\frac{1}{2}{ P}_k^2-\frac{1}{2} B_k^2
+A_0\partial_k P_k
\label{ssm50}
\ee
Next a canonical transformation is invoked. There are two possibilities
(assigning different signatures for the dual projection) which translate
from the old set $(P_k, A_k)$ to the new ones $( A_k^1, A_k^2)$.
It is, however, important to recall that the Maxwell theory has a Gauss constraint
that is implemented by the Lagrange multiplier $A_0$. The new variables
are chosen in two different ways which solve this constraint and implement
distinct signatures to the dual projection as,
\ba
P_k&\rightarrow& B_k^2\,\,\,\,;\,\,\,\, A_k\rightarrow  A_k^1\nonumber\\
 P_k&\rightarrow&  B_k^1\,\,\,\,;\,\,\,\, A_k\rightarrow  A_k^2\label{ssm60}
\ea
It is now simple to show that, in terms of the new variables, the
original Maxwell action takes the form,
\be
S_\pm={1\over 2}\int d^4x\; \left(\pm {\dot A}_k^\alpha
\epsilon^{\alpha\beta} B_k^\beta - B_k^\alpha B_k^\alpha\right)
\label{ssm70}
\ee
It is duality symmetric under the full $SO(2)$.

Let us next introduce the proper and improper $O(2)$ rotation matrices as $R^+(\theta)$ and
$R^-(\varphi)$ with determinant $+1$ and $-1$, respectively,

\ba
R^+\left(\theta\right)
=
\left(\begin{array}{cc}
{\cos\theta} & {\sin\theta} \\
{-\sin\theta} &{\cos\theta} \end{array}\right)
\label{ssmatrix1}
\ea
\ba
R^-\left(\varphi\right)
=
\left(\begin{array}{cc}
{\sin\varphi} & {\cos\varphi} \\
{\cos\varphi} &{-\sin\varphi} \end{array}\right)
\label{ssmatrix}
\ea
Note that the matrix that corresponds to improper rotations, $R^-(\varphi)$
switches the actions $S_+$ and $S_-$ into one another.

Using the basic brackets following from the canonical transformations or
from the symplectic structure of the theory,
\be
\label{ssgirotti}
\Big [A^i_\alpha(x),
\epsilon^{jkl}\partial^k A^l_\beta(y)\Big]=\pm i\delta^{ij}
\epsilon_{\alpha\beta} \delta({\bf{x}}-{\bf{y}})
\ee
we can verify that the generators of the $SO(2)$ rotations are given by
a Chern-Simon like structure,
\be
\label{ssjohn}
Q^{(\pm)}=\mp\frac{1}{2}\int d^3x\,\, {{A}}_k^\alpha\,\,{{B}}_k^\alpha
\ee
so that finite transformations are given by,
\be
\label{ssgt}
A_k^\alpha\rightarrow {A'}_k^\alpha=e^{-iQ\theta}{ {A}}_k^\alpha
e^{iQ\theta} 
\ee
Let us stress on the fact that there are 
two distinct structures for the duality symmetric
actions. These must correspond to the opposite aspects of some
symmetry. By looking at the equations of
motion obtained from (\ref{ssm70}),
\be
 {\dot A}_k^\alpha =
 \pm\epsilon^{\alpha\beta} \nabla_k \times  A_k^\beta
\label{ssm80}
\ee 
it is possible to
verify that these are just the 
self and anti-self dual solutions,
\be
F_{\mu\nu}^\alpha=\pm\epsilon^{\alpha\beta}\mbox{}^* F_{\mu\nu}^\beta
\,\,;\,\,\mbox{}^* F_{\mu\nu}^\beta=\frac{1}{2}\epsilon_{\mu\nu\rho\lambda}F^{\rho
\lambda}_\beta
\label{ssm90}
\ee
obtained by setting $A_0^\alpha=0$.
It may be observed that the opposite aspects
of the dual symmetry are contained in the internal space.

To close our arguments, let us now comment on another
property, which is related to the existence of two distinct
actions (\ref{ssm70}), by replacing (\ref{ssm40}) with a new set of
transformations,
\ba
E_\alpha&\rightarrow& R^-_{\alpha\beta}(\varphi)E_\beta\nonumber\\
B_\alpha&\rightarrow& R^-_{\alpha\beta}(\varphi)B_\beta
\label{ss110b}
\ea
Notice that these transformations preserve the invariance of
the hamiltonian following from either $S_+$ or $S_-$. The
kinetic terms in the action change signatures so that $S_+$ swaps to
$S_-$.
The discretised version of (\ref{ss110b}) is obtained by setting 
$\varphi =0$,
\ba
E_\alpha&\rightarrow& \sigma_1^{\alpha\beta}E_\beta\nonumber\\
B_\alpha&\rightarrow& \sigma_1^{\alpha\beta}B_\beta
\label{ss110a}
\ea
It is precisely the $\sigma_1$ matrix that reflects the proper into
improper rotations,
\be
\label{sseverton}
R^+(\theta) \sigma_1=R^-(\theta)
\ee
which illuminates the reason behind the swapping of the actions in
this example.

\subsection{The Scalar Theory in 1+1 Dimensions}

The ideas developed in the previous section are now implemented and
elaborated in $1+1$
dimensions. In particular we show that two distinct dual projections are
possible in this case, leading to either $Z_2$ or SO(2) group of dualities.
Notice first that in D=2 there is no photon and
the Maxwell theory trivialises so that the electromagnetic field can be
identified with a scalar field. Thus all the results presented here can be
regarded as equally valid for the ``photon" field.  There is no Gauss
constraint however so that we are free to choose any operator  in the dual projection.
Our computations will be presented in a very suggestive
notation which  illuminates the Maxwellian nature of the problem.
The action for the free massless scalar field is given by,
\be
\label{ssw10}
S=\frac{1}{2}\int d^2x\; \Big(\partial_\mu\phi\Big)^2
\ee
and the equation of motion reads,
\be
\label{ssw20}
\ddot\phi-\phi ''=0
\ee
where the dot and the prime denote derivatives with respect to time and
space components, respectively. Introduce a 
change of variables using electromagnetic symbols,
\be
E=\dot\phi\,\,\,\,;\,\,\,\, B=\phi '
\label{ssw30}
\ee
Obviously, $E$ and $B$ are not independent but constrained by the
identity, 
\be
\label{ssw40}
E'-\dot B=0
\ee
In these variables the equation of motion and the action are expressed
as,
\ba
\label{ssw50}
&\mbox{}&\dot E-B'=0\nonumber\\
&\mbox{}&S=\frac{1}{2}\int d^2x\; \Big(E^2-B^2\Big)
\ea
so that the transformations,
\be
\label{ssw60}
E\rightarrow \pm B\,\,\,\,;\,\,\,\, B\rightarrow \pm E
\ee
display a duality between the equation of motion and the `Bianchi'-like
identity (\ref{ssw40}) but the action changes its signature.
Note that there is a relative change in the signatures
of the duality transformations (\ref{ssw60}) with respect to the true
electromagnetic duality (\ref{ssm40}), arising
basically from dimensional considerations. This symmetry coresponds to the
improper group of rotations.

To illuminate the close connection with the Maxwell formulation, we
introduce covariant and contravariant vectors with a Minkowskian metric
$g_{00}=-\, g_{11}=1$, 
\be
F_\mu=\partial_\mu\phi\,\,\,;\,\,\, F^\mu=\partial^\mu\phi
\label{ssw60a}
\ee
whose components are just the `electric' and
`magnetic' fields defined earlier,
\be
F_\mu=\Big(E, B\Big)\,\,\,\,;\,\,\,\, F^\mu=\Big(E,- B\Big)
\label{ssw60b}
\ee
Likewise, with the convention $\epsilon_{01}=1$, the dual field is defined,
\ba
\mbox{}^* F_\mu&=&\epsilon_{\mu\nu}\partial^\nu\phi =\epsilon_{\mu\nu}F^\nu 
\nonumber\\
&=& \Big(-B, -E\Big)
\label{ssw60c}
\ea
The equation of motion and the `Bianchi' identity are now expressed by
typical electrodynamical relations,
\ba
\partial_\mu F^\mu&=&0\nonumber\\
\partial_\mu \mbox{}^* F^\mu&=&0
\label{ssw60d}
\ea

To expose a
duality symmetric action, the basic principle of our approach is adopted.
We convert the original second order form
(\ref{ssw50}) to its first order version displaying the Sp(2) symmetry and then invoke a canonical
transformation to provide an internal index.   
An auxiliary field is therefore introduced at the first step,
\be
\label{ssw70}
{\cal L}= PE-\frac{1}{2}P^2-\frac{1}{2}B^2
\ee
where $E$ and $B$ have already been defined. The following canonical transformation,

\ba
B &\to & \partial\left(\Phi^{(+)} + \Phi^{(-)}\right)\nonumber\\
P &\to & \partial\left(\Phi^{(+)} - \Phi^{(-)}\right)
\ea
leads to an action with fields taking values in the internal space

\be
S=\int d^2x \left[ \left(\partial\Phi^{(+)}\dot\Phi^{(+)} -\partial\Phi^{(-)}
\dot\Phi^{(-)}\right) +  \left(\partial\Phi^{(+)}\dot\Phi^{(-)} -\partial\Phi^{(-)}
\dot\Phi^{(+)}\right) - \left(\partial\Phi^{(\alpha)}\partial\Phi^{(\alpha)}\right)\right]
\ee
As discussed previously, due to the absence of a true Gauss law in this case,
we are free of any imposition regarding the choice of the operator $\partial$
in the dual projection.  To display
this arbitrariness, we choose, for each group of symmetry transformation, 
\be
\label{sp20}
\partial = \cases{\partial_x,&leading to $Z_2$\cr
	\sqrt{-\partial^2_x},&leading to $SO(2)$.\cr}
\ee

The first choice is traditional.  The odd parity of the operator diagonalizes
the action by eliminating the second term in the sympletic sector. 
The resulting actions, 
\ba
\label{sp10}
S &=& S_+ + S_-\nonumber\\
S_{\pm} &=& \int d^2 x \left(\dot\Phi^\pm \partial_x\Phi^\pm -
\partial_x\Phi^\pm \partial_x\Phi^\pm\right)
\ea
correspond to the well known right and left
chiral boson theories\cite{FJ}.

To examine the symmetry content it is possible to recast (\ref{sp10})
in a very suggestive form,
\ba
\label{ssw90} 
S_\pm&=&  \frac{1}{2} \int d^2 x \left[\pm
{\partial_x\Psi}^\alpha\sigma_1^{\alpha\beta}\dot\Psi^\beta -
{\partial_x\Psi}^\alpha {\partial_x\Psi}^\alpha \right] 
\nonumber\\
&=&  \frac{1}{2} \int d^2 x \left[\pm
B_\alpha\sigma_1^{\alpha\beta}E_\beta-B_\alpha^2\right] 
\ea
where $\Psi^\pm = \Phi^+ \pm \Phi^-$. In the second line the
action is expressed in terms of the electromagnetic variables.
This action is
duality symmetric under the transformations of the basic scalar fields,
\be
\label{ssw100}
\Psi_\alpha\rightarrow\sigma^{\alpha\beta}_1\Psi_\beta
\ee
which, in the notation of $E$ and $B$, is given by,
\ba
\label{ssw110}
B_\alpha &\rightarrow &\sigma^{\alpha\beta}_1B_\beta\nonumber\\
E_\alpha &\rightarrow &\sigma^{\alpha\beta}_1E_\beta
\ea
It is quite interesting to observe that, contrary to the 4D electromagnetic theory,  
the transformation matrix in the $O(2)$ internal space of potentials is not
the epsilon, but rather a $\sigma_1$ Pauli matrix. 
This result is in  agreement with that found from general algebraic
arguments \cite{DGHT} which stated that for $D=4k-2$ dimensions
there is a discrete $\sigma_1$ symmetry. Observe that 
(\ref{ssw110}) is a manifestation of the
original duality (\ref{ssw60}) which was also effected by the same operation.
It is important to stress that the above transformation is only
implementable at the discrete level. Moreover, since it is not connected
to the identity, there is no generator for it. In this sense it is observed that duality
symmetry is not defined in these twice odd dimensions.
To complete the picture, we also
mention that the following rotation,
\be
\Psi_\alpha\rightarrow \epsilon_{\alpha\beta}\Psi_\beta
\label{ssw120}
\ee
interchanges the actions (\ref{ssw90}),
\be
S_+\leftrightarrow S_-
\label{ssw130}
\ee
Thus, except for a rearrangement of the the matrices generating the
various transformations, most features of the electromagnetic example are
perfectly retained. The crucial
point of departure is that now all these transformations are only discrete.

The second choice in (\ref{sp20}) is new and unexpected since it does not
fit into the known dimensional classification.  It leads to a continous SO(2)
duality transformation, characteristic of the 4k dimensional spacetimes,
instead of the discrete $Z_2$.  The resulting action is,

\be
\label{sp30}
S=\int d^2x \left[ \left(\partial\Phi^{(+)}\dot\Phi^{(-)} -\partial\Phi^{(-)}
\dot\Phi^{(+)}\right) - \left(\partial\Phi^{(\alpha)}\partial\Phi^{(\alpha)}\right)\right]
\ee
which is appropriate for the even-parity dual projection.
It is easy to obtain the basic symplectic brackets from here as,

\be
\label{sp35}
\left\{\Phi^\alpha(x) , \partial \Phi^\beta(y)\right\}=\mp \epsilon^{\alpha\beta}
\delta(x-y)
\ee

Now observe
that the action (\ref{sp30}) 
is manifestly invariant under the continuous duality
transformations,
\be
\label{sp40}
\Phi^\alpha\rightarrow R^+_{\alpha\beta}\Phi^\beta
\ee
where $R^+_{\alpha\beta}$ is the usual $SO(2)$ rotation matrix (\ref{ssmatrix1}).
The generator of the infinitesimal symmetry transformation is given by,
\be
\label{sp50}
Q^\pm =\mp\frac{1}{2}\int d^2 x \Phi^\alpha \partial\Phi^\alpha
\ee
while the transformation by a finite angle $\theta$ is generated by,
\ba
\Phi^\alpha \rightarrow \tilde\Phi^\alpha & =& e^{-i\theta Q} \Phi^\alpha e^{i\theta Q}
\nonumber\\ 
&=& R_{\alpha\beta}^+(\theta)\Phi^\beta
\label{sp60}
\ea

We therefore conclude that the scalar theory in two dimensions manifests all features
of duality symmetry pertaining to either twice odd or twice even dimensions.
It, therefore, goes beyond the results found by the algebraic approach.

\section{Dual Projection in (2+1) dimensions}

As we have stressed, conventional group theoretical arguments
fail to discuss duality in odd dimensional theories.  This is
possible in our approach.  As an example we consider the (2+1) dimensional
Maxwell theory. We
shall see that the solution of the Gauss constraint leads naturally to
a differential operator to be used as the dual projector.
In subsection 4.1 the canonical transformation in the dual
projection involves an even-parity operation.  The resulting
action displays a
continous SO(2) group of symmetry transformations.  This is a new
result that could not be disclosed by algebraic methods.  However,
since in three spacetime dimensions vector fields are duality related
to scalars where there is no Gauss law restriction, an odd-parity kernel
also exists.
The complete diagonalization of the electromagnetic action into
two distinct type actions, that would be a prototype of 
chiral bosons in three dimensions, cannot be done in the coordinate space. 
On the other hand, if momentum space approach of the dual-projection is adopted,
such a structure is then shown to exist if a special combination
of the Fourier modes is considered.  This issue will be studied in subsection 4.2.

\subsection{Even Parity Projection}

To begin with we see that the solution of the Gauss constraint
that takes proper care of the spatial indices must involve a canonical scalar field,

\be
\label{q10}
\pi_k \sim \epsilon_{km}\partial_m\phi
\ee
The problem here is that, in contrast to the even dimensional cases, parity is not
a good property to look for in the generalized curl operator ($\epsilon\partial$).
However, exploiting the property,

\be\label{q20}
\nabla^2 = \left(\epsilon_{km}\partial_m\right)\left(\epsilon_{kn}\partial_n\right)
\ee
we may find an even solution for the dual projection by the following
canonical transformations,

\ba
\label{q30}
\pi_k &=& \eta \;\epsilon_{km} \partial_m\left(\phi^+ -\phi^-\right)\nonumber\\
A_k &=& \frac{\epsilon_{km}\partial_m}{\sqrt{-\nabla^2}}\left(\phi^+ +\phi^-\right)
\ea
where $\eta$ gives the signature of the dual projection and $\sqrt{-\nabla^2}$
is included for dimensional reasons.  Substituting these  into the Maxwell
action (\ref{a60}) we obtain,

\be
\label{q40}
S_\eta =\int d^3 x \left(\eta\dot\phi^\alpha \epsilon^{\alpha\beta}
B^\beta - B^\alpha B^\alpha\right)
\ee
where $B^\alpha$ is a shorthand for

\be
\label{q50}
B^\alpha =\sqrt{-\nabla^2} \; \phi^\alpha \;\;\; ; \;\;\;\phi^\alpha=\phi^+;\phi^-
\ee
Not surprisingly, the resulting action is explicitly duality invariant displaying
a continuous $SO(2)$ symmetry. 

Let us next examine the symmetry contents revealed
in (\ref{q40}).  These actions 
are manifestly invariant under the continuous SO(2)
transformations,
\be
\label{qe81}
\phi_\alpha\rightarrow R^+_{\alpha\beta}\phi_\beta
\ee
where $R^+_{\alpha\beta}$ is the proper $SO(2)$ rotation matrix 
(\ref{ssmatrix1}).
The generator of the infinitesimal symmetry transformation is given by the Chern-Simons form,
\be
\label{qe82}
Q_\eta =\frac{\eta}{2}\int d^3 x \,\phi_\alpha B_\alpha
\ee
and the finite transformations (\ref{qe81}) are generated as,
\ba
\phi_\alpha \rightarrow \tilde\phi_\alpha & =& e^{-i\theta Q} \phi_\alpha e^{i\theta Q}
\nonumber\\ 
&=& R_{\alpha\beta}^+(\theta)\phi_\beta
\label{qe83}
\ea
This result comes by using the basic symplectic brackets obtained
from (\ref{q40}),
\be
\label{qe84}
\{\phi_\alpha(\vec x) , B_\beta(\vec y)\}=
\eta \,\epsilon_{\alpha\beta}\,\delta(\vec x - \vec y)
\ee
This is the parallel of the usual constructions 
done in the 4D Maxwell theory and its D=4k extensions to
induce a duality symmetry in the action.

It is interesting to check the dual projection procedure when applied to
a (2+1) dimensional scalar field theory.  Recall that a vector field in
3D has only one degree of freedom and spin zero.  The vector and the
scalar fields are related
by the dualization,

\be
\label{q60}
\mbox{}^* F_{\mu\nu} = \epsilon_{\mu\nu\lambda}\partial^\lambda \phi
\ee
A simple analysis shows that under this transformation, $B\to\dot\phi$ and
$E_k\to\epsilon_{km}\partial_m\phi$ meaning that there is an exchange of
the potential and sympletic sectors between the models.  Clearly the
Maxwell Gauss law is automatically satisfied in the scalar representation, 

\be
\label{q70}
{\bf\nabla} .{\bf E }= 0  \stackrel{*}{\longrightarrow} 
\partial_k\left(\epsilon_{km}\partial_m\phi\right)=0
\ee
showing that in the dual point of view the gauge constraint has no
dynamical consequences.  It is now easy to apply the dual projection
to the conventional scalar action, written in a first order form, 

\be
S=\int dx \left[ \pi \dot\phi - \frac 12 \pi^2 -
\frac 12 \left(\partial_i\phi\right)^2\right]
\ee
by performing the following
canonical transformations,

\ba
\label{q80}
\phi &=& \phi_+ + \phi_-\nonumber\\
\pi &=& \eta\,\sqrt{-\nabla^2}\left(\phi_+ - \phi_-\right)
\ea
The 
resulting action reproduces the result (\ref{q40}),
as it should.  This also shows the equivalence of these theories in the
context of dual projection.

\subsection{Odd Parity Projection}

To disclose an odd parity projection we note the fundamental difference from the two 
dimensional case. In the latter, there is only one space dimension and hence it is
straightforward to define the odd projection in the coordinate space. For any
space dimension greater than one, this projection becomes ambiguous
in the coordinate space.
The standard way to bypass this problem is to go over to the momentum space.
Let us therefore introduce a two-dimensional basis, $\left\{\hat e_a(k,x), 
\; a=1,2\right\}$, with $(k,x)$ being conjugate variables and the
orthonormalization condition given as,

\be
\label{210}
\int dx \;\hat e_a(k,x)\hat e_b(k',x)=\delta_{ab}\,\delta(k,k')
\ee
We choose the vectors in the basis to be eigenvectors of the Laplacian,
$\nabla^2=  \,\partial \cdot \partial$,

\be
\label{220}
\nabla^2 \hat e_a(k,x)= -\,\omega^2(k)    \hat e_a(k,x)
\ee
The action of $\partial$ over the $\hat e_a(k,x)$ basis is

\be
\label{230}
\partial \, \hat e_a(k,x)=\omega(k) M_{ab}\,\hat e_b(k,x)
\ee
that together with definition (\ref{220}) gives
\footnote{Here we use the matricial notation where $\left(\widetilde M\right)_{ab} = M_{ba}$.}

\be
\label{240}
\widetilde M\cdot M= - I
\ee
Let us use this basis to represent the elementary fields.  Since 
the Maxwell theory here is equivalent to a spin zero
scalar, it suffices to analyze this last case.
The canonical scalar and its conjugate momentum have
the following expansion,

\begin{eqnarray}
\label{250}
\Phi(x)&=&\int dk\; q_a(k)\;\hat e_a(k,x)\nonumber\\
\Pi(x) &=& \int dk\; p_a(k)\; \hat e_a(k,x)
\end{eqnarray}
with $q_a$ and $p_a$ being the expansion coefficients. It leads to a representation
of the action as a two-dimensional oscillator.
The phase-space is four-dimensional,
representing two degrees of freedom {\it per mode},

\be
\label{260}
S=\int dk \left\{p_a\dot q_a - \frac 12 p_a p_a -\frac {\omega^2}{2} q_a q_a\right\}
\ee
Let us now consider the following canonical transformation,

\begin{eqnarray}
\label{270}
p_a(k) &=& \omega(k)\;\epsilon_{ab}\left(\varphi_b^{(+)} - \varphi_b^{(-)}\right)\nonumber\\
q_a(k) &=& \left(\varphi_b^{(+)} + \varphi_b^{(-)}\right)
\ea
such that (\ref{260}) gets diagonalized,

\be
S=S_+ + S_-
\ee
where,

\be
\label{280}
S_\pm =\int dk \omega(k)\left(\pm\dot q_a \epsilon_{ab} q_b - \omega(k) q_a q_a \right)
\ee

This action displays the $Z_2$ symmetry since, under the transformation $\varphi_a^\alpha
\rightarrow \sigma_1^{\alpha\beta}\varphi_a^\beta$, the two pieces $S_+$ and $S_-$ are
swapped. Hence the theory shows both $SO(2)$ and $Z_2$ symmetries, depending upon the
nature of the transformation.

It may be useful to point out that the actions $S_\pm$
are the analogues actions for chiral bosons.
Each of these actions characterises
one degree of freedom {\it per mode} in phase space
or half degree of freedom in configuration space that represent a chiral scalar.  Since the
Maxwell theory is equivalent to a scalar field theory in three dimensional spacetime,
this is valid for the photon field as well.  This shows that it is indeed possible to obtain a
phase space reduced, diagonal selfdual solution for the Maxwell field in this odd dimensional
spacetime.

\section{Conclusions}

In this paper we have developed a new technique for obtaining
duality symmetric actions.  Different aspects of duality symmetry were discussed.
Our technique was
based on an operation which was termed as a dual projection. Since the
analysis was always carried out for first order systems an equivalence
between the lagrangian and hamiltonian approaches was possible. Indeed the
dual projection entailed a change of variables which was a canonical transformation 
in the phase space. The analysis was completely general, which required an appropriate
definition of the functional curl used for solving the Gauss law constraint.
The conventional results for the construction of duality symmetric actions
in any even dimension was contained in a single expression. Zooming in on the
particular even dimension, either $4k$ or $4k-2$, immediately displayed the
relevant $SO(2)$ or $Z_2$ symmetry, respectively. This sharp line
of distinction was related to the parity of the functional curl. Our analysis, however, went 
beyond. Using the same techniques it was possible to discuss the property of
duality symmetry in odd dimensions, which were not analysed earlier.
Since the definition of parity of the functional
curl was problematic in these dimensions, it was not straightforward to give 
a general discussion for arbitrary odd dimensions, as was the case for the even
dimensions. Hence we elaborated our methods by concentrating on the three 
dimensional Maxwell theory. 

Historically, the study of duality symmetry began by considering the symmetry
among the electric and magnetic fields. This led to an invariance of the 
equations of motion but not of the actions. However it was felt that
these were composites and one should study the symmetry properties
in the context of potentials which were regarded as the basic entities.
This was achieved by introducing an internal
space. The new duality symmetry involving the potentials preserved
the invariance of the corresponding actions. Two distinct classes of symmetries,
pertaining to either $4k$ or $4k-2$ dimensions were found. Now the potentials can
also be regarded as composites that were defined in terms of their
fourier modes. Pursuing this line of research  and studying the symmetries 
in terms of these modes we\cite{CW} were able to show that, suitably interpreted,
the distinction among the duality groups could be obiliterated. In other
words, in all even dimensions, both $SO(2)$ and $Z_2$ symmetry groups could coexist.

The present work goes a step further. We show that it is possible to obtain duality
symmetric actions for odd dimensions also. At the level of potentials,
the duality symmetry in a three dimensional Maxwell theory was shown to
posses the $SO(2)$ symmetry. Furthermore, by
passing to the Fourier modes the $Z_2$ symmetry was also revealed.
\vspace{1.0cm}

{\bf Acknowledgements} CW would like to thank the S.N. Bose Centre for Basic
Sciences for the invitation, kind hospitality and financial support during
his stay here. This author is partially supported by CNPq, FUJB and FAPERJ,
Brasilian scientific agencies.

\end{document}